\documentclass[showpacs,amsmath,amssymb,aps,10pt,reprint,superscriptaddress,prb]{revtex4-1}
\usepackage{graphicx}
\usepackage{bm}
\usepackage[breaklinks=true,colorlinks=true,linkcolor=blue,urlcolor=blue,citecolor=blue]{hyperref}
\usepackage{graphics}
\usepackage{dcolumn}
\usepackage{amsmath,amssymb}
\usepackage{mathdots}
\usepackage{natbib}
\usepackage{soul,color}
\usepackage{epstopdf}
\usepackage[note-name]{notes2bib}

\begin{document}

\title{Influence of annealing on the electrical resistance of YBCO single crystals}

\author{R. V. Vovk}
\author{G. Ya. Khadzhai}
\author{Z. F. Nazyrov}
\author{S.\,N. Kamchatnaya}
    \affiliation{Physics Department, V. Karazin Kharkiv National University, 61077 Kharkiv, Ukraine}
\author{O.~V.~Dobrovolskiy}
    \affiliation{Physikalisches Institut, Goethe University, 60438 Frankfurt am Main, Germany}
    \affiliation{Physics Department, V. Karazin Kharkiv National University, 61077 Kharkiv, Ukraine}
\author{A. Feher}
    \affiliation{Pavol Josef \u{S}af\'{a}rik University,\\
        Park Angelinum 9, 04154 Ko\u{s}ice,\\
        Slovakia}

\begin{abstract}
The effect of annealing on the basal-plane electrical resistivity of the YBa$_2$Cu$_3$O$_{7-\delta}$ single crystals is studied in a broad range of oxygen contents. Within the framework of s-d scattering of electrons by phonons, an increase in the oxygen deficit index, $\delta$, leads to a significant increase in the Debye temperature, $\theta$, which is associated with the isotropization of the phonon spectrum as the concentration of oxygen vacancies increases. Near the optimal doping, the role of the paraconductivity becomes crucial, whereas its contribution decreases with increasing $\delta$. At large values of $\delta$ some deviations from the s-d model of electron scattering by phonons are observed at room temperature, while no paraproductivity is observed. In the superconducting transition region, a 2D-3D crossover is observed, which shifts in the direction of $T_c$ with increasing $\delta$. The estimate for the transverse coherence length is about $1$\,\AA.
\end{abstract}
\maketitle

\section{Introduction}

Investigations of relaxation processes in non-stoichiometric superconducting cuprates \cite{Jor90pcs,Vov14ltp,Sad00prb,Sol16phb} belong to one of the most important research lines in the contemporary physics of high-temperature superconductivity. In particular, in the compound YBa$_2$Cu$_{3}$O$_{7-\delta}$, due to the presence of the labile oxygen, such processes can easily be induced by application of high pressure \cite{Sad00prb,Sol16phb,Sol16cap}, an abrupt temperature change \cite{Jor90pcs,Vov14ltp}, as well as result in consequence of long storage or aging \cite{Mar95apl,Vov14jms1}. This is accompanied by a substantial modification of the structure and the topology of the ensemble of defects \cite{Vov15jms}. The electrical transport characteristics are modified \cite{Kir93prb,Vov09jac,Vov11jac} as well. This represents an important tool to examine numerous theoretical models \cite{Bab99prb,Vov15pcs} and to search for high-$T_c$ compounds with robust technological characteristics \cite{Sol16prb}.

\enlargethispage{2\baselineskip}
Despite a huge number of works \cite{Jor90pcs,Vov14ltp,Sad00prb,Sol16phb,Sol16cap,Mar95apl,Vov14jms1,Vov15jms,Kir93prb,Vov09jac} devoted to these issues, a lot of details relating to the peculiarities of the realization of the non-equilibrium state in high-$T_c$ compounds of various composition remain uncertain so far. Of crucial importance are studies of the peculiar scattering mechanisms for normal and fluctuation-induced charge carriers \cite{Vov14cap,Vov14apa,Vov14ssc}. According to the contemporary views \cite{Ash11snm}, it is this question which may shed light on the microscopic nature of high-$T_c$ superconductivity, whose nature remains unclarified despite of a thirty-year-long history of theoretical and experimental investigations.
\begin{table*}[t!]
    \centering
    \begin{tabular}{|l|l|l|l|l|l|l|l|l|}
    \hline
    Sample &    $T_c$, K    &  $R_0$\,$m\Omega$m & $\theta$, K & $C_3$, m$\Omega$m & $b_1$, (m$\Omega$)$^{-1}$ & $T_1$, K & $B_2$   & $T_2$, K \\
    \hline
    s1 & 91.738 & 2.8 & 28 & 47.12 & 4$\times10^{-6}$ & 1091.5 & - & -\\
    \hline
    s2 & 90.96-90.50 & 63.65 & 372 & 643 & 7.1$\times10^{-6}$ & 609 & - & - \\
    \hline
    s3 & 88.80-87.12 & 69.60 & 368 & 652.5 & 7.1$\times10^{-6}$ & 615 & - & - \\
    \hline
    s4 & 88.39-87.12 & 79.35 & 347 & 703 & 1.24$\times10^{-5}$ & 557 & - & - \\
    \hline
    s5 & 78.515 & 8.55 & 380.8 & 1263 & - & - & -3 & 820\\
    \hline
    s6 & 57.79 & 100.7 & 802 & 5550 & - & - &-8.8 & 1220\\
    \hline
    s7 & 46-44 & 1285.5 & 1060.5 & 20100 & - & - & 1.55 & 1295\\
    \hline
    \end{tabular}
    \caption{\label{t1} Fitting parameters for $\rho_{ab}(T)$ by Eqs. (\ref{e1}) -- (\ref{e3}) of the YBa$_2$Cu$_3$O$_{7-\delta}$ single crystals.}
\end{table*}

Earlier in was shown (see, e.g. Refs. \cite{Mak00ufn,Vov12fnt,Vov17phb,Col65jap}) that the normal-state basal-plane electrical resistance of YBa$_2$Cu$_3$O$_{7-\delta}$ single crystals, $\rho_{nab}(T)$ can be with great accuracy fitted by the Bloch-Gr\"uneisen expression \cite{Mak00ufn,Kor15ltp} which describes electron scattering on phonons and defects. In this case the dependence $d\rho_{nab}(T)/dT$ demonstrates a not high, smeared maximum at $T_{mBG} \approx 0.35 \theta$ ($\theta$ is the Debye temperature) \cite{Vov12fnt,Vov17phb,Col65jap}, i.e. the dependence $\rho_{nab}(T)$ exhibits has a point of inflection.
At temperatures $T > \theta$, according to the  the Bloch-Gr\"uneisen expression, $\rho_{nab}(T)$ tends to a linear dependence with increasing temperature.
As the temperature is decreasing, $\rho_{nab}(T)$ deviates down from the high-temperature extrapolation $\rho\propto T$, that is associated with a passage from elastic to inelastic scattering. This deviation can also be related to the fluctuation paraconductivity \cite{Vov11jms} which is due to the presence of a pseudogap in YBa$_2$Cu$_3$O$_{7-\delta}$ demonstrates an exponential temperature variation.
The transition into the superconducting state causes the appearance of another, high and sharp maximum in $d\rho_{nab}(T)/dT$. If $T_{mBG} \leq T_c$, then the phonon maximum in $d\rho_{nab}(T)/dT$ is absent.
It should be emphasized that (i) phonon scattering always takes plays and (ii) the resistivity values for YBa$_2$Cu$_3$O$_{7-\delta}$ single crystals are high; they correspond to metallic systems with a pseudogap, such as amorphous allows, quasicrystals, dichalcogenides of transition metals and so on \cite{Mot90boo,All80boo}. Finally, (iii) in such systems a ``saturation'' of the resistivity is often observed, that is a deviation of the resistivity down from its high-temperature extrapolation $\rho \propto T$ with increasing temperature \cite{Gan13boo,San84pra}.
\enlargethispage{2\baselineskip}

In our previous work \cite{Kha18pcs} we showed that the temperature dependence of the basal-plane normal-state electrical resistance of optimally doped YBa$_2$Cu$_3$O$_{7-\delta}$ single crystals (with $T_c\simeq 92$\,K) can be with great accuracy approximated within the framework of the model of s-d electron-phonon scattering. Here, we approximate the dependence $\rho_{ab}(T)$ of YBa$_2$Cu$_3$O$_{7-\delta}$ single crystals to the Bloch-Gr\"uneisen expression with an account for the fluctuation conductivity and the resistivity ``saturation''. This allows us to deduce the effect of room-temperature annealing on the approximation parameters and the superconducting characteristics of YBa$_2$Cu$_3$O$_{7-\delta}$ single crystals with $T_c$ ranging between $44$ and $92$\,K.

\section{Experimental}
The YBa$_2$Cu$_3$O$_{7-\delta}$ single crystals were grown by the solution-melt technique in a gold crucible in the temperature range from $850^\circ$ to $970^\circ$ as in Refs. \cite{Vov14ltp,Sol16phb,Vov14jms1}. The typical crystal dimensions were $2\times0.3\times0.02$\,mm$^3$. The smallest crystal size corresponds to the $c$-axis. To obtain samples with optimal oxygen content, $\delta \leq 0.1$, selected crystals were annealed in a oxygen flow at $400^\circ$C for five days. To reduce the oxygen content, the samples were annealed in an oxygen flow at a higher temperature for three to five days. The effect of room-temperature annealing on the resistive properties was studied according to the procedure described in Ref. \cite{Vov14ltp}.
\enlargethispage{2\baselineskip}

The electrical contacts were formed by silver conductors attached to the crystal surface using a silver paste. Resistance measurements were done in the standard 4-probe geometry at a dc current of $1$\,mA. The temperature was measured by a thermocouple while the measurements were performed in a temperature sweep mode with a rate of $0.1$\,K/min near the superconducting transition temperature $T_c$ and $1$\,K/min at close-to-room temperatures.

In what follows we discuss the data for seven YBa$_2$Cu$_3$O$_{7-\delta}$ single crystals, referred to as samples s1 to s7, whose parameters are reported in Table \ref{t1} and explained in the next section.

\section{Results and discussion}
\subsection{Normal resistance and excess conductivity}
\begin{figure}[b!]
    \centering
    \includegraphics[width=0.8\linewidth]{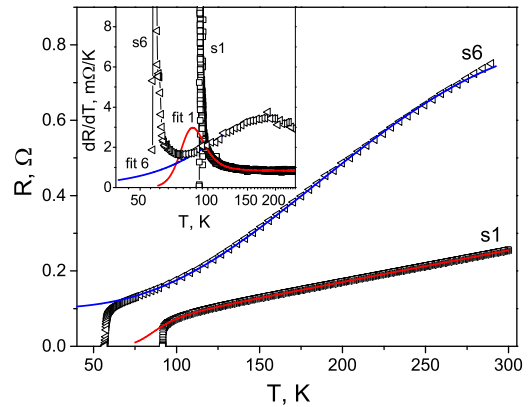}
    \caption{Temperature dependences of the electrical resistance of the YBa$_2$Cu$_3$O$_{7-\delta}$ samples s1 with $T_c = 91.738$\,K and s6 with $T_c = 57.79$\,K. Symbols: experiment. Lines: fit for s1 to Eqs. (\ref{e1}) and (\ref{e2}); fit for s6 to Eqs. (\ref{e2}) and (\ref{e3}). Inset: temperature dependences of the respective derivatives $d\rho_{ab}(T)/dT$: squares --- s1 with $T_c=91.738$\,K, as deduced from the experimental data, fit 1 --- calculated by Eqs. (\ref{e1}) and (\ref{e2}), triangles --- s6 with $T_c=57.79$\,K, as deduced from the experimental data, fit 6 --- calculated by Eqs. (\ref{e2}) and (\ref{e3}).}
    \label{f1}
\end{figure}

Figure \ref{f1} displays the experimental temperature dependences (symbols) of the electrical resistivity $\rho_{ab}(T)$ of the optimally doped (sample 1 with $T_c =91.738$\,K) and an annealed (sample 6 with $T_c =57.79$\,K) YBa$_2$Cu$_3$O$_{7-\delta}$ single crystal. The other curves are qualitatively similar: $\rho_{ab}(T)$ demonstrates a metallic temperature behavior. For the state with $T_c$ between 92\,K and 88\,K the dependence $\rho_{ab}(T)$ had to be approximated to the following expression
\begin{equation}
\label{e1}
    \rho_{nab}(T) = \frac{1}{\frac{1}{\rho_0 + \rho_{ph}} + b_1(e^{T_1/T} - 1)}.
\end{equation}
Here
\begin{equation}
\label{e2}
    \rho_{ph}(T) = C_3\left(\frac{T}{\theta}\right)^3\int_{0}^{\theta/T}\frac{e^x x^3 dx}{(e^x - 1)^2}.
\end{equation}
\enlargethispage{2\baselineskip}

In Eq. (\ref{e1}) $\rho_0$ is the residual resistivity, Eq. (\ref{e2}) is the Bloch-Gr\"uneisen expression, and the exponential term in Es. (\ref{e1}) describes the paraconductivity \cite{Vov11jms}, whose contribution is pronounced near $T_c$, that is at low temperatures.

For the state with $T_c$ between $80$\,K and $46$\,K the dependence $\rho_{ab}(T)$ was fitted to the expression
\begin{equation}
\label{e3}
    \rho_{nab}(T) = [\rho_0 + \rho_{ph}](1 + B_2e^{-T_2/T}).
\end{equation}
The exponential term in Eq. \ref{e3} describes the ``saturation'' of the resistivity \cite{Gan13boo,Boi17mre,Vov17ssc} that is, it chiefly contributes in the high-temperature region.

The fitting parameters to Eqs. (\ref{e1}) --- (\ref{e3}) were determined by least mean squares. The approximation error does not exceed $0.5\%$. The fitting parameters are reported in Table \ref{t1}.

The respective derivatives $d\rho_{nab}(T)/dT$ are plotted in the inset to Fig. \ref{f1}. For the state with $T_c =91.738$\,K (curve 1a) the high-temperature maximum in $d\rho_{nab}(T)/dT$ is in the superconducting region; this is why it is not observed experimentally. For the state with $T_c =57.79$\,K the high-temperature maximum in $d\rho_{nab}(T)/dT$ is observed near $185$\,K.


The increase of the oxygen index $\delta$ leads to an increase of the number of oxygen vacancies and their accumulation, that leads to a growth of the electrical resistance $R_0$, refer to Table \ref{t1}. However, for $T_c = 78.515$\,K a sudden drop of $R_0$ takes place, followed by a new increase. This behavior can be associated with a diffusion coalescence of the accumulated oxygen vacancies \cite{Vov17ssc}, which leads to the appearance of the single vacancy cluster and to the observed decrease in $R_0$. With a further increase of $\delta$ new clusters may appear so that $R_0$ increases once again.

The Debye temperature for the optimally doped single crystal is significantly lower that for the underdoped one \cite{Ans88etp}. In Ref. \cite{Kho83fnt} the temperature dependence of the specific heat of YBa$_2$Cu$_3$O$_{7-\delta}$ was described within the framework of a model accounting for transverse lattice oscillations along ($\theta_1\approx 90$\,K) and perpendicular to ($\theta_2\approx 850$\,K) the $c$-axis as well as longitudinal lattice oscillations with $\theta_3\approx 295$\,K. The values of $\theta$ deduced from you measurements are in line with the data of Ref. \cite{Kho83fnt} and attest to that in the optimally doped sample the charge carriers are primarily scattered on shear oscillations of the layers, while as $\delta$ increases, the interlayer oscillation become play the primarily role in the charge scattering. The reason for this is in the isotropization of the phonon spectrum with a gradual increase of the defect concentration.

The phonon scattering coefficient $C_3$ monotonically decreases with increasing $T_c$, that is as the lattice perfectness is improved. This change of $C_3$ can be associated with the deformation of the phonon spectrum of the sample in the presence of defects (see, e.g., \cite{Kag71etp,Azh03pla}, in our case of oxygen vacancies.

\begin{figure}[t!]
    \centering
    \includegraphics[width=0.8\linewidth]{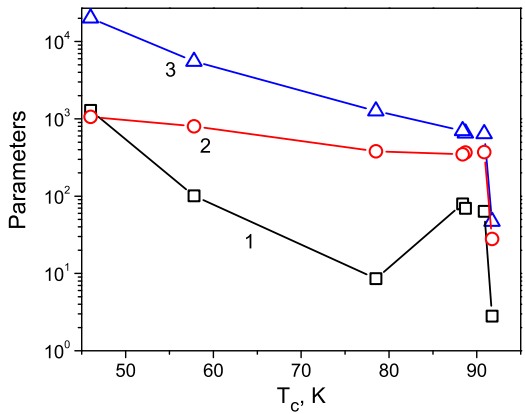}
    \caption{Dependences of the parameters of charge carriers scattering on phonons and defects on the superconducting transition temperature $T_c$: $1$ --- $R_0$,\,m$\Omega$, $2$ --- $\theta$,\,K, $3$ --- $C_3$,\,m$\Omega$.}
    \label{f2}
\end{figure}

The dependence of the parameters for charge scattering on phonons and defects on the superconducting transition temperature $T_c$ are shown in Fig. \ref{f2}.

\begin{figure}[t!]
    \centering
    \includegraphics[width=0.85\linewidth]{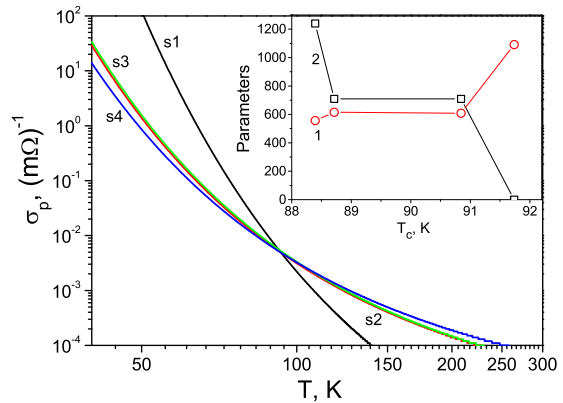}
    \caption{Temperature dependences of the paraconductivity $\sigma_p = b_1[\exp(T_1/T) -1]$ for samples s1 to s4.
    Inset: Dependences of the parameters $1$ --- $b_1$,\,m$\Omega^{-1}$ and $2$ --- $T_1$,\,K.}
    \label{f3}
\end{figure}

The temperature behavior of the exponential term $\sigma_p = b_1[\exp(T_1/T) - 1]$ in Eq. (\ref{e1}) describing the paraconductivity \cite{Vov11jms} is shown in Fig. \ref{f3}. One sees that all $\sigma_p(T)$ intersect at an almost the same point. Accordingly, at $T\geq 94$\,K $\sigma_p(T)$ increases with decreasing $T_c$, whereas for $T\leq 92$\,K $\sigma_p(T)$ decreases with decreasing $T_c$ (that is with increasing $\delta$).

With a further decrease of $T_c$ no paraconductivity is observed on the background of phonon scattering --- see Table \ref{t1}. Since the paraconductivity contributes strongly at low temperatures, where its contribution is decreasing with increasing $\delta$, one can assume that the paraconductivity is suppressed by defects, primarily by oxygen vacancies.

It should be noted the metallic systems with pseudogap can have a temperature dependence of the resistance which is described by the Bloch-Gr\"uneisen expression, see e.g., Refs. \cite{Kag71etp,Azh03pla}. The pseudogap value primarily depends on the material composition \cite{All80boo}, while its temperature dependence is weak and it is chiefly associated with the thermal expansion. For this reason the pseudogap associated with temperature can appear in the case when the pseudogap is a precursor of the superconducting transition and it turns into the superconducting gap as the temperature is decreasing \cite{Ais70psj}. Then, the paraconductivity can be considered as a contribution of occasionally appeared Cooper pairs with a coupling energy $kT_1 \sim 0.1$\,eV to the conductivity. The high-temperature term in Eq. (\ref{e3}) can be associated with various mechanisms and hence, it can have different forms \cite{Gan13boo,San84pra,Boi17mre,Zhu89boo,All80boo,Mor78cry,Keb89prb}.

\subsection{Superconducting transition}
\begin{figure}[t!]
    \centering
    \includegraphics[width=0.77\linewidth]{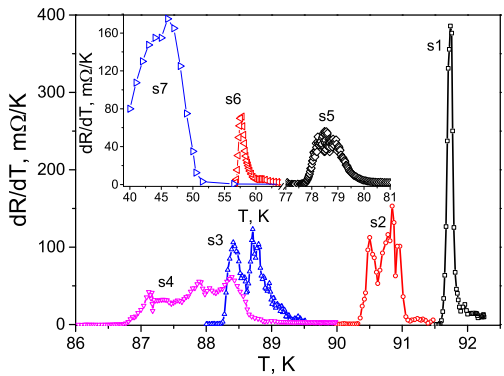}
    \caption{Derivatives $d\rho_{ab}(T)/dT$ for samples s1 to s7 with various oxygen concentrations.}
    \label{f4}
\end{figure}

The superconducting transition causes a decrease of the electrical resistance of the sample in a narrow temperature range. The superconducting temperature is usually determined at the point of the maximum in the derivation $d\rho{ab}(T)/dT$, whose width corresponds to the width of the superconducting transition. Figure \ref{f4} presents these maxima for different oxygen concentrations. One sees that for the optimally doped sample one narrow symmetric maximum is observed, while small variation in the oxygen content lead to the appearance of neighboring maxima, whose magnitude is decreasing with increasing $\delta$. This accompanied by an increase of their total width (refer to the main panel of Fig. \ref{f4}). For large $\delta$ one asymmetric maximum is observed. Its height increases with increasing $\delta$.

The presence of several maxima in $d\rho{ab}(T)/dT$ (or an asymmetric maximum) attests to the existence of macroscopic superconducting regions in the sample with different, but rather close $T_c$, that is regions with different $\delta$ values. In this case, vanish of the electrical resistance can be associated with the formation of a cluster of regions with the same $T_c$, which spreads over the entire sample. This $T_c$ corresponds to the most low-temperature (for a given $\delta$ value) maximum in $d\rho_{ab}(T)/dT$. Regions with lower $T_c$ can not be observed in resistance measurements. An oxygen exchange takes place between the different regions (of the diffusion coalescence type \cite{Vov17ssc}). This can lead to an equalization of $\delta$, as observed at large values of $\delta$, please refer to the inset of Fig. \ref{f1}. The inhomogeneous oxygen distribution in the sample is likely caused by the impurities and/or structural imperfectness such as, for instance, twins.

\subsection{2D-3D crossover}

In Ref. \cite{Ais70psj} for the basal-plane conductivity at temperatures right below the superconducting transition the following expression has been obtained for the Aslamazov-Larkin fluctuation conductivity \cite{Lar09boo}
\begin{equation}
\label{e4}
    \sigma_{ab}^{AL} = \frac{e^2}{16d}\frac{1}{\sqrt{\varepsilon(\varepsilon + T)}},
\end{equation}
where $d = 11.7$\,\AA~is the interlayer distance \cite{Fri89prb},
$$
\varepsilon = \frac{T-T_c}{T_c}\ll 1, \qquad r = 4(_\downarrow c^\uparrow 2(0)))/d^\uparrow2.
$$

In Eq. (\ref{e4}) at $\varepsilon \ll r$ (3D regime) $\sigma_{ab}^{AL} \propto (\varepsilon r)^{-1/2}$, whereas at $\varepsilon \gg r$ (2D regime) $\sigma_{ab}^{AL} \propto (\varepsilon)^{-1}$. This is known is a crossover from the high-temperature 2D regime to the low-temperature 3D regime, whereby the condition $\varepsilon_{cross} \sim r$ determines the crossover region. It should be noted that in Eq. (\ref{e4}) the parameter $r$ is the only characteristics of the sample.
\begin{figure}[t!]
    \centering
    \includegraphics[width=0.75\linewidth]{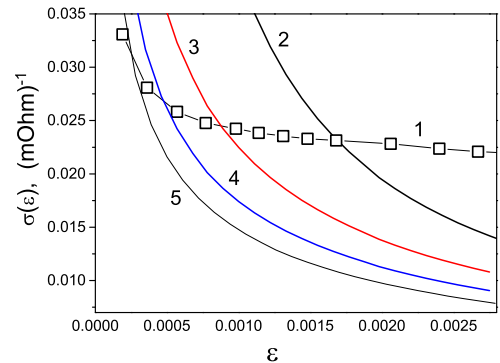}
    \caption{Dependences $\sigma(\varepsilon)$ near the superconducting transition. $1$ --- experiment for the initial state ($T_c = 91.738$). Solid lines --- calculation by Eq. (\ref{e4}) for $2$ --- $r =0$, $3$ --- $r=0.002$, $4$ --- $r = 0.005$, and $5$ --- $r = 0.006$.}
    \label{f5}
\end{figure}

Figure \ref{f5} displays the experimental dependences $\sigma(\varepsilon) = 1/\rho_{ab}(\varepsilon)$ near the superconducting transition and the dependences $\sigma_{ab}^{AL}(\varepsilon)$ calculated by Eq. (\ref{e4}) for different $r$ values. One sees the qualitative agreement of the experimental data and calculation results at $\varepsilon \leq 0.0017$ and $r > 0.006$. This means that $\varepsilon < r$ that is in this case $\xi_c(T) < d$ and the most low-temperature experimental values of $\sigma_{ab}(T)$ are stipulated the 3D motion of the charge carriers. Other specific mechanisms of quasiparticle scattering \cite{Apa02prb65,Vov03prb,Ada94ltp,Vov03prl,Cur11prb} may also play a role.

The crossover condition \cite{Sol02ltp2} reads $\xi_c(T_{cross}) \approx d$, that is $\xi_c(0) \approx d(r_{cross})^{1/2}$. This yields $\xi_c(0)\approx1\,\AA$ that agrees with the results of Ref. \cite{Sol02ltp3}.

With increasing $\delta$ the experimental data agree with the calculation results only for $\varepsilon >r$, that is the most low-temperature experimental values $\sigma_{ab}(T)$ are stipulated by the 2D motion of the charge carriers. This means that with increasing $\delta$ the crossover region tends to approach $T_c$. One should note that a comparison of $\sigma(\varepsilon)$ with $\sigma_{ab}^{AL}(\varepsilon)$ can be made only in the case of well-defined $T_c$ that is in the case of one maximum in $d\rho_{ab}(T)/dT$.

\section{Conclusion}
The following conclusions can be made from our study. (i) The temperature dependence of the normal-state basal-plane electrical resistance of YBa$_2$Cu$_3$O$_{7-\delta}$ single crystals near the optimal doping level can be with great accuracy described within the framework of the model of s-d-scattering of electrons on phonons with an account for the paraconductivity whose contribution exponentially increases with decreasing temperature. (ii) At large $\delta$, deviations from the model of s-d-scattering of electrons on phonons are observed at close-to-room temperatures. No paraconductivity is observed. (iii) Maxima in $d\rho_{ab}(T)/dT$ in the superconducting transition region attest to an inhomogeneous oxygen distribution in the sample, most likely due to defects. (iv) The behavior of the conductivity near the superconducting transition agrees well with the Aslamazov-Larkin model, a 2D-3D crossover is observed, and the transverse coherence length amounts to about $1\,\AA$.

The research leading to these results has received funding from the European UnionТs Horizon 2020 research and innovation program under Marie Sklodowska-Curie Grant Agreement No. 644348 (MagIC).


\end{document}